\documentclass[aps,pre,twocolumn,groupedaddress,showpacs]{revtex4}

\usepackage{graphicx}
\usepackage{dcolumn}
\usepackage{bm}
\usepackage{amssymb}

\newcommand{\correct}[1]{#1}
\newcommand{\newcorrect}[1]{#1}

\begin{document}

\title{Detection of synchronization from univariate data using wavelet
transform}

\author{Alexander~E.~Hramov}
\email{aeh@nonlin.sgu.ru}
\author{Alexey~A.~Koronovskii}
\email{alkor@nonlin.sgu.ru} \affiliation{Faculty of Nonlinear
Processes, Saratov State University, Astrakhanskaya, 83, Saratov,
410012, Russia}
\author{Vladimir I. Ponomarenko}
\email{vip@sgu.ru}
\author{Mikhail D. Prokhorov}
\affiliation{Saratov Department of the Institute of
RadioEngineering and Electronics of Russian Academy of Sciences,
Zelyonaya, 38, Saratov, 410019, Russia}

\date{\today}

\begin{abstract}
A method is proposed for detecting from univariate data the
presence of synchronization of a self-sustained oscillator by
external driving with varying frequency. The method is based on
the analysis of difference between the oscillator instantaneous
phases calculated using continuous wavelet transform at time
moments shifted by a certain constant value relative to each
other. We apply our method to a driven asymmetric van der Pol
oscillator, experimental data from a driven electronic oscillator
with delayed feedback and human heartbeat time series. In the
latest case, the analysis of the heart rate variability data
reveals synchronous regimes between the respiration and slow
oscillations in blood pressure.
\end{abstract}


\maketitle

\vskip 0.5cm

\section{Introduction}
\label{Sct:Introduction}

Detecting regimes of synchronization between self-sustained
oscillators is a typical problem in studying their interaction.
Two types of interaction are generally recognized
\cite{Blekhman:1971_SynhroBookEngl, Blekhman:1988_SynchroBook,
Pikovsky:2002_SynhroBook, Boccaletti:2002_ChaosSynchro}. The first
one is a unidirectional coupling of oscillators. It can result in
synchronization of a self-sustained oscillator by an external
force. In this case the dynamics of the oscillator generating the
driving signal does not depend on the driven system behavior. The
second type is a mutual coupling of oscillators. In this case the
interaction can be more effective in one of the directions,
approaching in the limit to the first type, or can be equally
effective in both directions. In the event of mutual coupling,
synchronization is the result of the adjustment of rhythms of
interacting systems. To detect synchronization one can analyze the
ratio of instantaneous frequencies of interacting oscillators and
the dynamics of the generalized phase difference
\cite{Pikovsky:2002_SynhroBook}. As a quantitative characteristic
of synchronization one can use the phase synchronization index
\cite{Rosenblum:2001_HandbookBiologicalPhysics,
Meinecke:2005_Prosachivanie} or the measure of synchronization
\cite{Hramov:2004_Chaos, Aeh:2005_SpectralComponents}.

Synchronization of interacting systems including the chaotic ones
has been intensively studied in recent years. The main ideas in
this area have been introduced using standard models
\cite{Blekhman:1971_SynhroBookEngl, Blekhman:1988_SynchroBook,
Pecora:1990_ChaosSynchro, Pecora:1997_SynchroChaos,
Pikovsky:2000_SynchroReview, Boccaletti:2001_UnifingSynchro,
Pikovsky:2002_SynhroBook, Boccaletti:2002_ChaosSynchro,
Rulkov:1995_GeneralSynchro, Pyragas:1996_WeakAndStrongSynchro,
Hramov:2004_Chaos, Aeh:2005_SpectralComponents}). At present, more
attention is focused on application of the developed techniques to
living systems. In particular, much consideration is being given
to investigation of synchronization between different brain areas
\cite{Tass:1998_NeuroSynchro, Tass:2003_NeuroSynchro,
Meinecke:2005_Prosachivanie, Boccaletti:2005_Chaos_TSS} and to
studying synchronization in the human cardiorespiratory system
\cite{Schafer:1999_cardio, Stefanovska:2000_cardio_Physica_A,
Rzeczinski:2002_cardio, Prokhorov:2003_HumanSynchroPRE,
Hramov:2006_Prosachivanie}. Investigating such systems one usually
deals with the analysis of short time series heavily corrupted by
noise. In the presence of noise it is often difficult to detect
the transitions between synchronous and nonsynchronous regimes.
Besides, even in the region of synchronization a $2\pi$-phase
jumps in the temporal behavior of the generalized phase difference
can take place. Moreover, the interacting systems can have a set
of natural rhythms. That is why it is desirable to analyze
synchronization and phase locking at different time scales
\cite{Hramov:2004_Chaos, Hramov:2005_JETP, Aeh:2005_TSS:PhysicaD,
Boccaletti:2005_Chaos_TSS, Hramov:2005_Chaos_BWO}.

A striking example of interaction between various rhythms is the
operation of the human cardiovascular system (CVS). The main
rhythmic processes governing the cardiovascular dynamics are the
main heart rhythm, respiration, and the process of slow regulation
of blood pressure and heart rate having in humans the fundamental
frequency close to 0.1\,Hz \cite{Malpas:2002_cardio}. Owing to
interaction, these rhythms appear in various signals:
electrocardiogram (ECG), blood pressure, blood flow, and heart
rate variability (HRV) \cite{Stefanovska:2000_cardio}. Recently,
it has been found that the main rhythmic processes operating
within the CVS can be synchronized \cite{Schafer:1999_cardio,
Stefanovska:2000_cardio_Physica_A, Rzeczinski:2002_cardio,
Prokhorov:2003_HumanSynchroPRE}. It has been shown that the
systems generating the main heart rhythm and the rhythm associated
with slow oscillations in blood pressure can be regarded as
self-sustained oscillators, and that the respiration can be
regarded as an external forcing of these systems
\cite{Prokhorov:2003_HumanSynchroPRE, Rzeczinski:2002_cardio}.

Recently, we have proposed a method for detecting the presence of
synchronization of a self-sustained oscillator by external driving
with linearly varying frequency \cite{Hramov:2006_Prosachivanie}.
This method was based on a continuous wavelet transform of both the
signals of the self-sustained oscillator and external force.
However, in many applications the diagnostics of synchronization
from the analysis of univariate data is a more attractive problem
than the detection of synchronization from multivariate data. For
instance, the record of only a univariate signal may be available
for the analysis or simultaneous registration of different variables
may be rather difficult. In this paper we propose a method for
detection of synchronization from univariate data. However, a
necessary condition for application of our method is the presence of
a driving signal with varying frequency. For the mentioned above
cardiovascular system our method gives a possibility to detect
synchronization between its main rhythmic processes from the
analysis of the single heartbeat time series recorded under paced
respiration.

The paper is organized as follows. In Sec.~II we describe the method
for detecting synchronization from univariate data. In Sec.~III the
method is tested by applying it to numerical data produced by a driven
asymmetric van der Pol oscillator. In Sec.~IV the method is used for
detecting synchronization from experimental time series gained from a
driven electronic oscillator with delayed feedback. Section~V presents
the results of the method application to studying synchronization
between the rhythms of the cardiovascular system from the analysis of
the human heart rate variability data. In Sec.~VI we summarize our
results.

\section{Method description}
\label{Sct:Method}

Let us consider a self-sustained oscillator driven by external force
${\cal F}$ with varying frequency
\begin{equation}
 \dot{\textbf{x}}={\textbf{H}}(\textbf{x}) +
  \varepsilon{\cal F}(\Phi(t)),
 \label{eq:Generator}
\end{equation}
where $\textbf{H}$ is the operator of evolution, $\varepsilon$ is the
driving amplitude, and $\Phi(t)$ is the phase of the external force
defining the law of the driving frequency $\omega_d(t)$ variation:
\begin{equation}
 \omega_d(t)=\frac{d\Phi(t)}{dt}.
 \label{eq:FreqExtSignal}
\end{equation}
In the simplest case the external force is described by a harmonic
function ${\cal F}(\Phi(t))=\sin\Phi(t)$.

Assume that we have at the disposal a univariate time series
$x(t)$ characterizing the response of the
oscillator~(\ref{eq:Generator}) to the driving force $\cal F$. Let
us define from this time series the phase $\varphi_0(t)$ of
oscillations at the system~(\ref{eq:Generator}) basic frequency
$f_0$. The main idea of our approach for detecting synchronization
from univariate data is to consider the temporal behavior of the
difference between the oscillator instantaneous phases at the time
moments $t$ and $t+\tau$. We calculate the phase difference
\begin{equation}
\Delta\varphi_0(t)=\varphi_0(t+\tau)-\varphi_0(t),
 \label{eq:DeltaPhasa0}
\end{equation}
where $\tau$ is the time shift that can be varied in a wide range.
Note, that $\varphi_0(t)$ and $\varphi_0(t+\tau)$ are the phases of
the driven self-sustained oscillator corresponding to oscillations at
the first harmonic of the oscillator basic frequency $f_0$.

The variation of driving frequency is crucial for the proposed
method. Varying in time, the frequency of the external force
sequentially passes through the regions of synchronization of
different orders $1:1$, $2:1$, \dots, $n:1$,~\dots, $n:m$, \dots
($n,m=1,2,3, \dots$). Within the time intervals corresponding to
asynchronous dynamics the external signal practically has no
influence on the dynamics of the basic frequency $f_0$ in the
oscillator~(\ref{eq:Generator}) spectrum. Thus, the phase of
oscillator varies linearly outside the regions of synchronization,
$\varphi_0(t)=2\pi f_0 t + \bar\varphi$, where $\bar\varphi$ is
the initial phase. Then, from Eq.~(\ref{eq:DeltaPhasa0}) it
follows
\begin{equation}
 \Delta\varphi_0(t)=2\pi f_0\tau,
 \label{eq:DeltaPhasaAsinchro}
\end{equation}
i.e., the phase difference $\Delta\varphi_0(t)$ is constant within the
regions of asynchronous dynamics.

Another situation is observed in the vicinity of the time moments
$t_{ns}$ where the driving frequency $\omega_d(t)\approx (2\pi
n/m)f_0$ and $n:m$ synchronization takes place. For simplicity let
us consider the case of $1:1$ synchronization. In the
synchronization (Arnold) tongue the frequency of the
system~(\ref{eq:Generator}) nonautonomous oscillations is equal to
the frequency~(\ref{eq:FreqExtSignal}) of the external force and
the phase difference between the phase of the driven oscillator
$\varphi_{0}(t)$ and the phase $\Phi(t)$ of the external force,
$\Delta\tilde{\phi}(t)=\varphi_{0}(t)-\Phi(t)$, is governed in a
first approximation by the Adler equation \cite{Adler:1949}. It
follows from the Adler equation that in the region of $1:1$
synchronization the phase difference $\Delta\tilde{\phi}(t)$
varies by $\pi$.

Representing the driven oscillator phase as
$\varphi_{0}(t)=\Delta\tilde{\phi}(t)+\Phi(t)$, we obtain from
Eq.~(\ref{eq:DeltaPhasa0}):
\begin{equation}
\Delta\varphi_0(t)=\Phi(t+\tau)-\Phi(t)+\gamma,
 \label{eq:PhaseLocking2a}
\end{equation}
where
$\gamma=\Delta\tilde{\phi}(t+\tau)-\Delta\tilde{\phi}(t)\approx\rm
const$ is the correction of the phase difference that appears due to
synchronization of the system by external force. Expanding the phase
$\Phi(t+\tau)$ in a Taylor series we obtain
\begin{equation}
\Delta\varphi_0(t)=\gamma+\frac{d\Phi(t)}{dt}\tau+\frac12\frac{d^2\Phi(t)}{dt^2}\tau^2+\dots.
 \label{eq:PhaseLocking2b}
\end{equation}
Taking into account Eq.~(\ref{eq:FreqExtSignal}) we can rewrite
Eq.~(\ref{eq:PhaseLocking2b}) as
\begin{equation}
\Delta\varphi_0(t)=\gamma+\omega_d(t)\tau+\frac12\frac{d\omega_d(t)}{dt}\tau^2+\dots.
 \label{eq:PhaseLocking2c}
\end{equation}
Thus, the behavior of the phase difference~(\ref{eq:DeltaPhasa0}) is
defined by the law of the driving frequency $\omega_d(t)$ variation.

For the linear variation of the driving frequency,
$\omega_d(t)=\alpha+\beta t$, from Eq.~(\ref{eq:PhaseLocking2c}) it
follows
\begin{equation}
\Delta\varphi_0(t)=\gamma+\alpha\tau+\beta\tau^2/2+\tau\beta t.
 \label{eq:PhaseLocking2d}
\end{equation}
Consequently, in the region of synchronization the phase
difference varies linearly in time, $\Delta\varphi_0(t)\sim t$. In
the case of the nonlinear variation of $\omega_d(t)$, the dynamics
of $\Delta\varphi_0(t)$ is more complicated. However, if
$\omega_d(t)$ varies in a monotone way and the time of its passing
through the synchronization tongue is small, one can neglect the
high-order terms of the expansion and consider the law of
$\Delta\varphi_0(t)$ variation as the linear one. We will show
below that this assumption holds true for many applications.

The absolute value of the change in the phase difference
$\Delta\varphi_0(t)$ within the synchronization region can be
estimated using Eq.~(\ref{eq:PhaseLocking2c}):
$$
\Delta\varphi_s=\Delta\varphi_0(t_2)-\Delta\varphi_0(t_1)=(\omega_d(t_2)-\omega_d(t_1))\tau+
$$
\begin{equation}
+\left(\left.\frac{d\omega_d(t)}{dt}\right|_{t=t_2}-\left.\frac{d\omega_d(t)}{dt}\right|_{t=t_1}\right)\frac{\tau^2}{2}+\dots,
\label{eq:PhaseLocking2c1}
\end{equation}
where $t_1$ and $t_2$ are the time moments when the frequency of the
external force passes through, respectively, the low-frequency and
high-frequency boundaries of the synchronization tongue. Assuming that
the rate of $\omega_d(t)$ variation is slow, we can neglect the terms
containing the derivatives of $\omega_d(t)$ and obtain
\begin{equation}
\Delta\varphi_s\approx\Delta\omega\tau,
\label{eq:PhaseLocking2g}
\end{equation}
where $\Delta\omega=\omega_d(t_2)-\omega_d(t_1)$ is the bandwidth of
synchronization.

The obtained estimation corresponds to the case of $1:1$
synchronization, characterized by equal values of the driving
frequency $f_d$ and the oscillator frequency $f_0$, $f_d/f_0=1$.
However, the considered approach can be easily extended to a more
complicated case of $n:m$ synchronization. In this case the change in
$\Delta\varphi_0(t)$ within the region of synchronization takes the
value
\begin{equation}
\Delta\varphi_s=\frac{m}{n}\Delta\omega\tau.
 \label{eq:PhaseLocking2f}
\end{equation}
Hence, the analysis of the phase difference~(\ref{eq:DeltaPhasa0})
behavior allows one to distinguish between the regimes of synchronous
and asynchronous dynamics of driven oscillator. The phase difference
$\Delta\varphi_0(t)$ is constant for the regions of asynchronous
dynamics and demonstrates monotone (often almost linear) variation
by the value $\Delta\varphi_s$ defined by
Eq.~(\ref{eq:PhaseLocking2f}) within the regions of synchronization.

To define the phase $\varphi_0(t)$ of oscillations at the basic
frequency we use the approach based on the continuous wavelet
transform \cite{Koronovskii:2004_JETPLettersEngl,
Hramov:2004_Chaos, Aeh:2005_TSS:PhysicaD,
Aeh:2005_SpectralComponents}.
\correct{It is significant, that the wavelet
transform~\cite{WaveletsInPhysics:2004, alkor:2003_WVTBookEng} is
the powerful tool for the analysis of nonlinear dynamical system
behavior. The continuous wavelet analysis has been applied in the
studies of phase synchronization of chaotic neural oscillations in
the brain \cite{Lachaux:1999, Lachaux:2000, Lachaux:2001,
Lachaux:2002_BrainCoherence, Quyen:2001_HTvsWVT},
electroencephalogram signals \cite{Quiroga:2002}, R--R intervals and
arterial blood pressure oscillations in brain injury
\cite{Turalska:2005}, chaotic laser
array~\cite{DeShazer:2001_WVT_LaserArray}. It has also been used to
detect the main frequency of the oscillations in nephron
autoregulation~\cite{Sosnovtseva:2002_Wvt} and coherence between
blood flow and skin temperature oscillations
\cite{BANDRIVSKYY:2004}.} \newcorrect{In these recent studies a
continuous wavelet transform with various mother wavelet functions
has been used for introducing the instantaneous phases of analyzed
signals. In particular, in Refs.\,\cite{Lachaux:2001, Quiroga:2002}
a comparison of Hilbert transform and wavelet method with the mother
Morlet wavelet has been carried out and good conformity between
these two methods has been shown for the analysis of neuronal
activity. It is important to note, that in all the above mentioned
studies the wavelet transform has been used for the analysis of
synchronization from bivariate data, when the generalized phase
difference $\Delta\varphi(t)$ of both analyzed rhythms was
investigated. The proposed method allows one to detect
synchronization from the analysis of only the one signal of the
oscillator response to the external force with monotonically varying
frequency. Taking into account the high efficiency of the analysis
of synchronization with the help of the continuous wavelet transform
using bivariate data, we will use the continuous wavelet transform
for determining the instantaneous phase of the analyzed univariate
signal.}

The continuous wavelet transform \cite{WaveletsInPhysics:2004,
alkor:2003_WVTBookEng} of the signal $x(t)$ is defined as
\begin{equation}
W(s,t_0)=\int_{-\infty}^{+\infty}x(t)\psi^*_{s,t_0}(t)\,dt,
\label{eq:WvtTrans}
\end{equation}
where $\psi_{s,t_0}(t)$ is the wavelet function related to the
mother wavelet $\psi_{0}(t)$ as
$\psi_{s,t_0}(t)=\left({1}/{\sqrt{s}}\right)\psi_0\left(({t-t_0})/{s}\right)$.
The time scale $s$ corresponds to the width of the wavelet
function, $t_0$ is the shift of the wavelet along the time axis,
and the asterisk denotes complex conjugation. It should be noted
that the wavelet analysis operates usually with the time scale $s$
instead of the frequency $f$, or the corresponding period $T=1/f$,
traditional for the Fourier transform.

The wavelet spectrum
\begin{equation}
W(s,t_0)=|W(s,t_0)|\exp[j\varphi_s(t_0)]
\label{eq:WVT_Phase}
\end{equation}
describes the system dynamics for every time scale $s$ at any time
moment $t_0$. The value of $|W(s,t_0)|$ determines the presence
and intensity of the time scale $s$ at the time moment  $t_0$. We
use the complex Morlet wavelet \cite{Grossman:1984_Morlet}
$\psi_0(\eta)=({1}/{\sqrt[4]{\pi}})\exp[j\sigma\eta]\exp\left[{-
\eta^2}/{2}\right]$ as the mother wavelet function. The choice of
the wavelet parameter $\sigma=2\pi$ provides the simple relation
$f\approx1/s$ between the frequency $f$ of the Fourier transform
and the time scale $s$ \cite{alkor:2003_WVTBookEng}.

\section{Method application to detecting synchronization in a driven
asymmetric van der Pol oscillator}
\label{Sct:Exp1_VDP}

\subsection{Model}
\label{SubSct:Exp1_VDP_A}

Let us consider the asymmetric van der Pol oscillator under external
force with linearly increasing frequency:
\begin{equation}
\ddot{x}-\left(1-\mu
x-x^2\right)\dot{x}+\Omega^2x=\varepsilon\sin\Phi(t),
\label{eq:VdP}
\end{equation}
where $\mu$ is the parameter characterizing the system asymmetry,
$\Omega=0.24\pi$ is the natural frequency, and $\varepsilon$ and
$\Phi(t)$ are, respectively, the amplitude and phase of the
external force. The phase $\Phi(t)=2\pi\left[(\alpha+\beta
t/T)\right]t$ defines the linear dependence of the driving
frequency $\omega_d(t)$ on time:
\begin{equation}
\omega_d(t)=\frac{d\Phi(t)}{dt}=2\pi\left[\alpha+2\beta
t/T\right], \label{eq:ExtSignal}
\end{equation}
where $\alpha=0.03$, $\beta=0.17$, and $T=1800$ is the maximal
time of computation. This system has been considered in
Ref.~\cite{Hramov:2006_Prosachivanie} as a model for studying
synchronization between the respiration, which can be regarded as
an external force, and the process of slow regulation of blood
pressure and heart rate, which can be treated as a self-sustained
oscillator. In the present paper we use this model system for
testing our new method of detecting synchronization from
univariate data. The chosen values of the model parameters provide
close correspondence of frequencies and the ways of the driving
frequency variation in the simulation and experimental study
described in Sec.~V. The parameter $\mu$ is chosen to be equal to
unity throughout this paper. In this case the phase portrait of
oscillations is asymmetric and the power spectrum contains both
odd and even harmonics of the basic frequency $f_0=0.0973$, as
well as the power spectrum of the low-frequency fluctuations of
blood pressure and heart rate \cite{Hramov:2006_Prosachivanie}.
Recall that the classical van der Pol oscillator with $\mu=0$ has
a symmetric phase portrait and its power spectrum exhibits only
odd harmonics of $f_0$. We calculate the time series of
nonautonomous asymmetric van der Pol oscillator~(\ref{eq:VdP}) at
$\varepsilon=0.2$ using a fourth-order Runge-Kutta method with the
integration step $\Delta t=0.01$.

\subsection{Results}
\label{SubSct:Exp1_VDP_B}

\begin{figure}
 \centerline{\scalebox{0.4}{\includegraphics{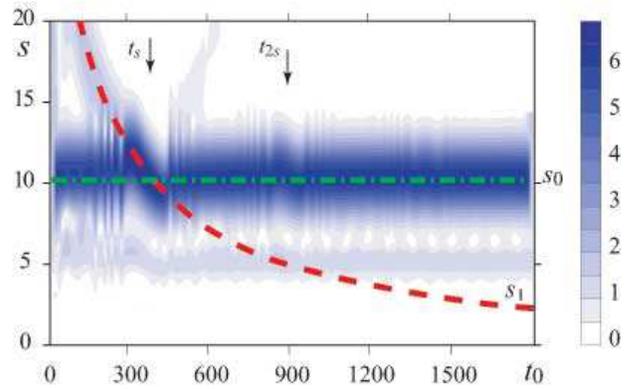}}}
\caption{(Color online) Shaded plot of the wavelet power spectrum
$|W(s,t_0)|$ for the signal generated by
oscillator~(\ref{eq:VdP}). Time is shown on the abscissa and time
scale is shown on the ordinate. The color intensity is
proportional to the absolute value of the wavelet transform
coefficients. The values of the coefficients are indicated by the
scale from the right side of the figure. \label{fig1}}
\end{figure}

Fig.~\ref{fig1} shows the amplitude spectrum $|W(s,t_0)|$ of the
wavelet transform for the signal of driven
oscillator~(\ref{eq:VdP}). The Morlet wavelet is used as the mother
wavelet function throughout the paper. The wavelet parameter is
chosen to be $\sigma=2\pi$, unless otherwise specified. The time
scale $s_0$ corresponding to the first harmonic of the oscillator
basic frequency $f_0$ is indicated in Fig.~\ref{fig1} by the
dot-and-dash line. The dashed line indicates the time scale $s_1$
corresponding to the linearly increasing driving frequency
$\omega_d(t)$. The analysis of the wavelet power spectrum reveals
the classical picture of oscillator frequency locking by the
external driving. As the result of this locking, the breaks appear
close to the time moments $t_s$ and $t_{2s}$ denoted by arrows, when
the driving frequency is close to the oscillator basic frequency
($\omega_d(t_s)\approx 2\pi f_0$) or to its second harmonic
($\omega_d(t_{2s})\approx 4\pi f_0$), respectively. These breaks
represent the entrainment of oscillator frequency and its harmonic
by external driving. If the detuning $\delta = (\omega_d-2\pi f_0)$
is great enough, the frequency of oscillations returns to the
oscillator basic frequency.

\begin{figure}
 \centerline{\scalebox{0.35}{\includegraphics{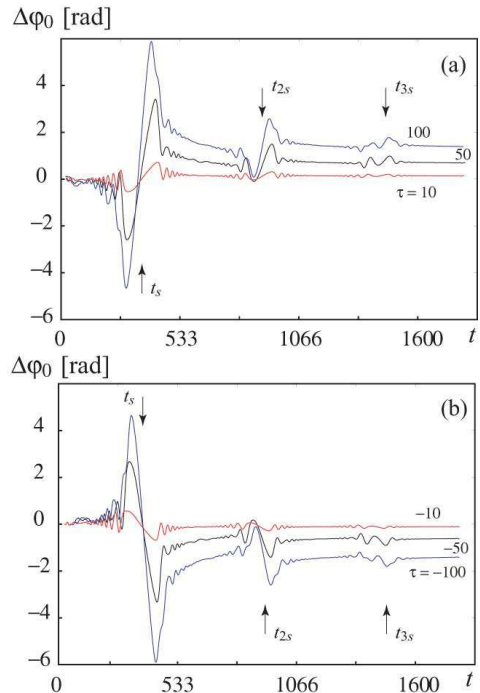}}}
\caption{(Color online) Phase differences $\Delta\varphi_0(t)$
(\ref{eq:DeltaPhasa0}) calculated at the time scale $s_0$
corresponding to the basic frequency $f_0=0.0973$ of the driven
asymmetric van der Pol oscillator~(\ref{eq:VdP}) for different
$\tau>0$ (a) and $\tau<0$ (b).
 \label{fig2}}
\end{figure}

The dynamics of the phase differences $\Delta\varphi_0(t)$
determined by Eq.~(\ref{eq:DeltaPhasa0}) is presented in
Fig.~\ref{fig2}a for different positive $\tau$ values. One can see
in the figure the regions where $\Delta\varphi_0(t)$ is almost
constant. These are the regions of asynchronous dynamics, when the
driving frequency is far from the oscillator basic frequency and
its harmonics. The regions of monotone increase of
$\Delta\varphi_0(t)$ are also well-pronounced in Fig.~\ref{fig2}a.
These are the regions of synchronization observed in the vicinity
of the time moments $t_{ns}$, when $\omega_d(t_{ns})\approx2\pi
nf_0$.

The proposed method offers several advantages over the method in
Ref.~\cite{Hramov:2006_Prosachivanie} based on the analysis of the
phase difference between the signals of oscillator and the external
force. First, the regions of $\Delta\varphi_0(t)$ monotone variation
corresponding to synchronous regimes are easily distinguished from
the regions of constant $\Delta\varphi_0(t)$ value corresponding to
asynchronous dynamics. Second, the new method is considerably more
sensitive than the previous one because the phase difference is
examined at the time scales having high amplitude in the wavelet
spectrum. In particular, the region of $3:1$ synchronization in the
vicinity of the time moment $t_{3s}$ denoted by arrow is clearly
identified in Fig.~\ref{fig2}. Third, the proposed method is
substantially simpler than the method of the phase difference
calculation along the scale varying in time
\cite{Hramov:2006_Prosachivanie}.

It follows from Eq.~(\ref{eq:PhaseLocking2c}) that in the region
of synchronization the change of the phase difference
$\Delta\varphi_0(t)$ increases with $\tau$ increasing. As the
result, the presence of interval of $\Delta\varphi_0(t)$ monotone
variation becomes more pronounced, Fig.~\ref{fig2}a. This feature
helps to detect the existence of synchronization especially in the
case of high-order synchronization and noise presence. However,
the accuracy of determining the boundaries of the region of
synchronization decreases as $\tau$ increases.

It should be noted that for negative $\tau$ values the monotone
reduction of the phase difference is observed in the region of
synchronization, Fig.~\ref{fig2}b. As it can be seen from
Fig.~\ref{fig2}b, the increase of $\tau$ by absolute value leads
to increase of $\Delta\varphi_0(t)$ variation in the region of
synchronization as well as in the case of positive $\tau$.

\subsection{Influence of noise and inaccuracy of the basic time scale
definition}
\label{SubSct:Exp1_VDP_C}

Experimental data, especially those obtained from living systems, are
always corrupted by noise. Besides, in many cases it is not possible
to define accurately the basic frequency of the system under
investigation. For example, interaction between the human
cardiovascular and respiratory systems and nonstationarity hampers
accurate estimation of natural frequencies for cardiovascular rhythms.
Therefore, the actual problem is to test the method efficiency for
detecting synchronization in the presence of additive noise and
inaccuracy of the basic frequencies estimation.

\begin{figure}
 \centerline{\scalebox{0.5}{\includegraphics{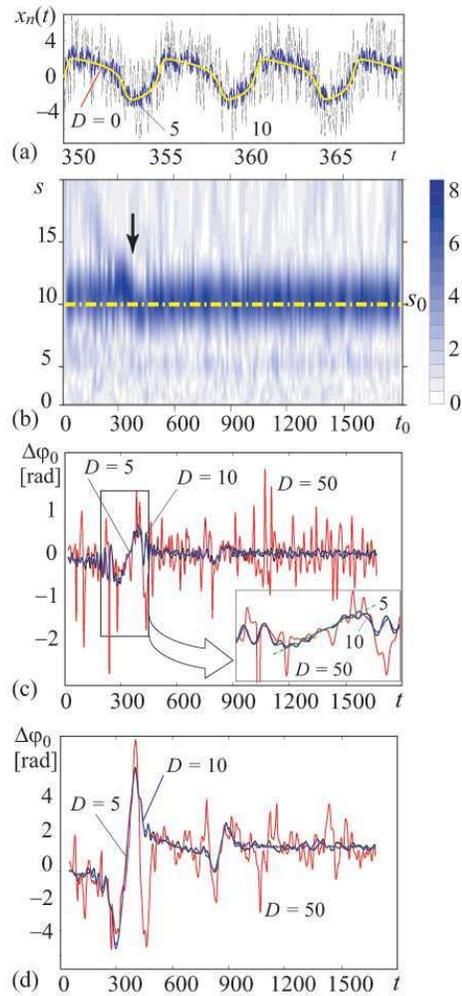}}}
\caption{(Color online) (a) Parts of the time series of the
signal~(\ref{eq:Noise+VdP}) for different intensities $D$ of
additive noise. (b) Wavelet power spectrum $|W(s,t_0)|$ of the
signal $x_n(t)$ at the noise intensity $D=10$. The dot-and-dash line
indicates the time scale $s_0$ corresponding to the oscillator basic
frequency $f_0$. (c, d) Phase differences $\Delta\varphi_0(t)$ for
different intensities $D$ of noise at $\tau=10$ (c) and $\tau=100$
(d). The inset in (c) is the enlarged fragment of the region of
$1:1$ synchronization.
 \label{fig3}}
\end{figure}

To analyze the influence of noise on the diagnostics of
synchronization we consider the signal
\begin{equation}
 x_n(t)=x(t)+D\zeta(t),
 \label{eq:Noise+VdP}
\end{equation}
where $x(t)$ is the signal of the asymmetric van der Pol oscillator
(\ref{eq:VdP}), $\zeta(t)$ is the additive noise with zero mean and
uniform distribution in the interval $[-0.5,\,0.5]$, and $D$ is the
intensity of noise. To simulate the noisy signal $\zeta(t)$ we use the
random-number generator described in
Ref.~\cite{NumericalRecipes:1997}.

Typical time series $x_n(t)$ generated by Eq.~(\ref{eq:Noise+VdP})
for different intensities of noise are presented in
Fig.~\ref{fig3}a for the region of $1:1$ synchronization. In spite
of the significant distortion of the signal by noise its wavelet
power spectrum, Fig.~\ref{fig3}b, still allows to reveal the main
features of the system dynamics. In particular, the dynamics of
the time scale $s_0$ and the effect of frequency entrainment in
the region of $1:1$ synchronization indicated by arrow are
recognized in Fig.~\ref{fig3}b. Hence, the use of the wavelet
transform for determining the phases of the signal and its
harmonics allows one to detect the regimes of synchronization from
noisy time series.

The phase differences $\Delta\varphi_0(t)$ calculated using
Eq.~(\ref{eq:DeltaPhasa0}) with $\tau=10$ are shown on
Fig.~\ref{fig3}c for different intensities $D$ of additive noise.
The dependence $\Delta\varphi_0(t)$ becomes more \correct{jagged} as
$D$ increases. However, for $D<10$ we can identify the regions where
the phase difference demonstrates near-monotone variation. In the
average this variation is about the same as in the case of noise
absence (see the inset in Fig.~\ref{fig3}c). Fig.~\ref{fig3}d shows
$\Delta\varphi_0(t)$ for $\tau=100$. In this case it is possible to
detect the presence of synchronization for significantly higher
levels of noise than in the case of small $\tau$. The reason is that
the value of $\Delta\varphi_s$~(\ref{eq:PhaseLocking2f}) increases
in the region of synchronization as the time shift $\tau$ increases,
whereas the amplitude of $\Delta\varphi_0(t)$ fluctuations caused by
noise does not depend on $\tau$. For very large intensities of noise
($D=50$ in Fig.~\ref{fig3}) the synchronous behavior is not so
clearly pronounced as at smaller $D$ values, but it should be
mentioned that in this case the power of noise exceeds the power of
the oscillator signal in several times.

\begin{figure}
 \centerline{\scalebox{0.43}{\includegraphics{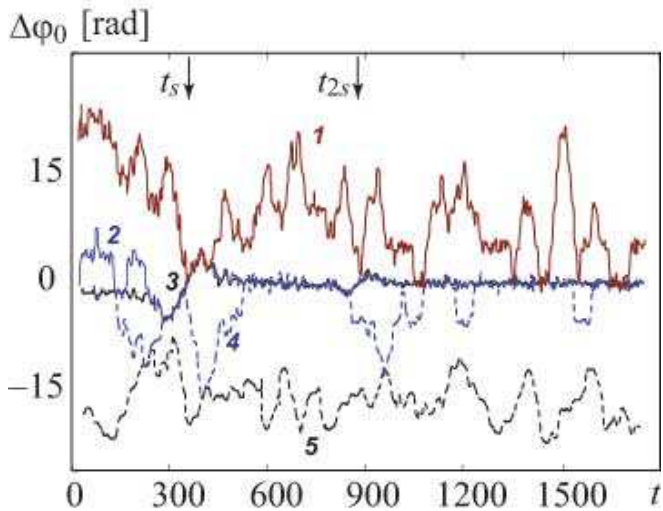}}}
\caption{(Color online) Phase differences $\Delta\varphi_0(t)$
calculated at the time scales $s_1=s_0+\Delta s$ for $\tau=100$
and $D=10$. The curve numbers correspond to the following time
scaled: (1)~$s_1=7.28<s_0$, (2)~$s_1=8.28<s_0$,
(3)~$s_1=s_0=10.28$, (4)~$s_1=12.28>s_0$, (5)~$s_1=15.28>s_0$.
 \label{fig4}}
\end{figure}

Let us consider the method efficiency in the case when the scale $s$
of observation differs from the time scale $s_0$ associated with the
oscillator basic frequency $f_0$. Fig.~\ref{fig4} illustrates the
behavior of the phase difference $\Delta\varphi_0(t)$ calculated for
the time series of Eq.~(\ref{eq:Noise+VdP}) at the time scales
$s_1=s_0+\Delta s$, where $\Delta s$ is the detuning of the scale
with respect to the basic scale $s_0\approx1/f_0=10.28$. It can be
seen from the figure that for $|\Delta s|<2.0$ the phase dynamics is
qualitatively similar to the case of accurate adjustment of the
scale $s$ to the basic scale $s_0$. At greater $\Delta s$ values the
phase difference demonstrates significant fluctuations impeding to
detect the epochs of $\Delta\varphi_0(t)$ monotone variation. Thus,
to detect synchronization using the proposed method one needs to
know only approximately the basic time scale $s_0$.

\section{Investigation of synchronization in a driven electronic
oscillator with delayed feedback}
\label{Sct:Exp2_Experiment}

\subsection{Experiment description}

\label{Sct:Exp2_Experiment_A}

\begin{figure*}
 \centerline{\scalebox{0.6}{\includegraphics{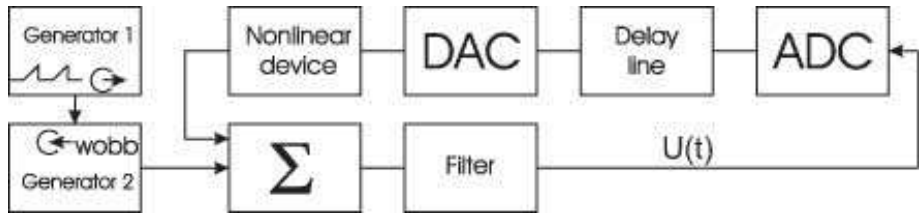}}}
\caption{Block diagram of the electronic oscillator with delayed
feedback driven by the signal with varying frequency.
 \label{fig_experiment_schema}}
\end{figure*}

We apply the method to experimental data gained from a driven
electronic oscillator with delayed feedback. A block diagram of
the experimental setup is shown in
Fig.~\ref{fig_experiment_schema}. The oscillator represents the
ring system composed of nonlinear, delay, and inertial elements.
The role of nonlinear element is played by an amplifier with the
quadratic transfer function. This nonlinear device is constructed
using bipolar transistors. The delay line is constructed using
digital elements. The inertial properties of the oscillator are
defined by a low-frequency first-order $RC$-filter. The analogue
and digital elements of the scheme are connected with the help of
analog-to-digital (ADC) and digital-to-analog converters (DAC). To
generate the driving signal we use the sine-wave generator~{\sl2}
whose frequency is modulated through the wobble input by the
signal of the sawtooth pulse generator~{\sl1}. The driving signal
is applied to the oscillator using the summator $\Sigma$. The
considered oscillator is governed by the first-order time-delay
differential equation
\begin{equation}
 RC \dot{U}(t)=-U(t)+F(U(t-d))+U_0\sin(2\pi f_{ext}(t)t),
 \label{eq:TimeDelaySystem}
\end{equation}
where $U(t)$ and $U(t-d)$ are the delay line input and output
voltages, respectively, $d$ is the delay time, $R$ and $C$ are the
resistance and capacitance of the filter elements, $F$ is the transfer
function of the nonlinear device, $U_0$ is the amplitude of the
driving signal, and $f_{ext}$ is the driving frequency. We record the
signal $U(t)$ using an analog-to-digital converter with the sampling
frequency $f=15$\,kHz at $d=1.5$\,ms and $RC=0.46$\,ms under the
following variation of the driving frequency
\begin{equation}
f_{ext}(t)=\nu\cdot10^{U_w(t)/2},
\label{Eq:Freq_Experiment}
\end{equation}
where $\nu=220\,$Hz and the control voltage $U_w(t)$ varies linearly
from 0\,V to 16\,V within 800\,ms providing $f_{ext}$ variation from
220\,Hz to 1000\,Hz. Under the chosen parameters the considered
oscillator demonstrates periodic oscillations with the period
$T=3.7$\,ms. Four experiments were carried out at different amplitudes
of the external driving equal to 0.5\,V, 1\,V, 1.5\,V, and 2\,V. The
amplitude of driven oscillation was about 3\,V.

\subsection{Results}
\label{Sct:Exp2_Experiment_B}

\begin{figure}
 \centerline{\scalebox{0.4}{\includegraphics{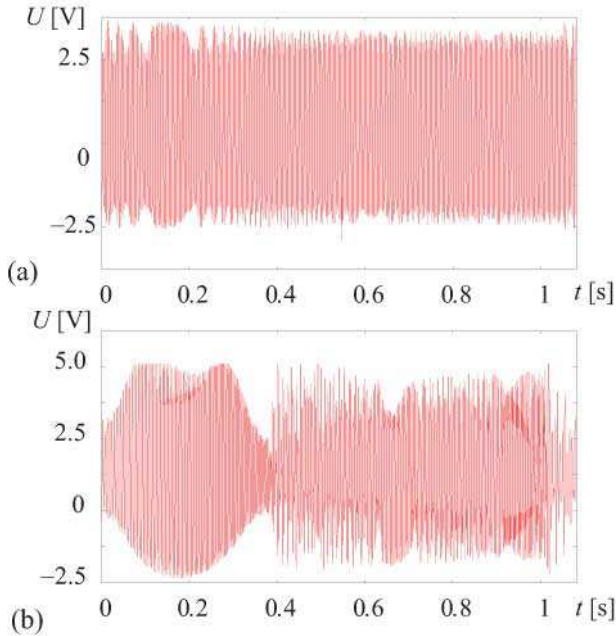}}}
\caption{(Color online) Time series of electronic oscillator with
delayed feedback under external driving with varying
frequency~(\ref{Eq:Freq_Experiment}) and the driving amplitude
$U_0=0.5$\,V~(a) and $U_0=2$\,V~(b).
 \label{fig_experiment_TimeSeries}}
\end{figure}

\begin{figure}
 \centerline{\scalebox{0.4}{\includegraphics{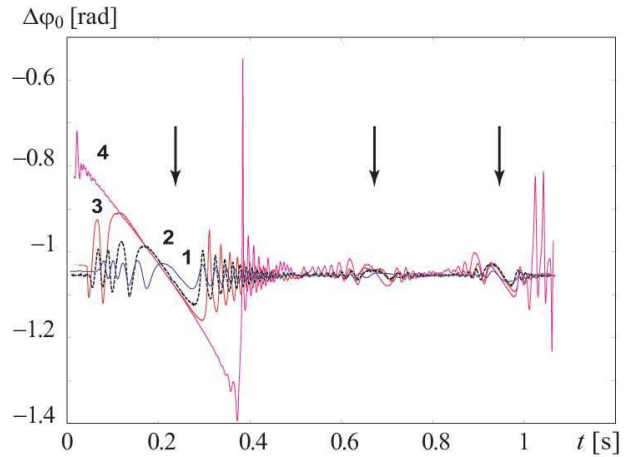}}}
\caption{(Color online) Phase differences $\Delta\varphi_0(t)$
(\ref{eq:DeltaPhasa0}) calculated at the time scale $s_0$
corresponding to the basic frequency $f_0=270$\,Hz of the driven
electronic oscillator with delayed feedback. The curve numbers
correspond to different amplitudes $U_0$ of the external force:
(1)~$U_0=0.5$\,V, (2)~$U_0=1$\,V, (3)~$U_0=1.5$\,V,
(4)~$U_0=2$\,V.
 \label{fig_experiment_Phases}}
\end{figure}

The experimental time series of the electronic oscillator with
delayed feedback driven by the external force with varying
frequency~(\ref{Eq:Freq_Experiment}) are depicted in
Fig.~\ref{fig_experiment_TimeSeries} for two values of the driving
amplitude. The results of investigation of the oscillator
synchronization by the external driving are presented in
Fig.~\ref{fig_experiment_Phases}. The phase differences
$\Delta\varphi_0(t)$ defined by Eq.~(\ref{eq:DeltaPhasa0}) are
calculated under different driving amplitudes $U_0$ for the time
shift $\tau=-0.66$\,ms. One can clearly identify in the figure the
regions of $\Delta\varphi_0(t)$ monotone variation corresponding
to the closeness of the driving frequency to the oscillator basic
frequency and its harmonics. These regions of synchronous dynamics
are indicated by arrows.

It is well seen from Fig.~\ref{fig_experiment_Phases} that the
interval of monotone variation of $\Delta\varphi_0(t)$ increases with
increasing amplitude of the driving force. This fact agrees well with
the known effect of extension of the region of synchronization with
increase in the amplitude of the external driving. Note, that in spite
of the nonlinear variation of the driving frequency, at small driving
amplitudes the phase difference $\Delta\varphi_0(t)$ varies almost
linearly in time within the synchronization tongue as it was discussed
in Sec.~\ref{Sct:Method}. For the large driving amplitude ($U_0=2$\,V)
the synchronization tongue is wide enough and the phase difference
behavior begins to depart from linearity. Nevertheless, the variation
of $\Delta\varphi_0(t)$ remains the monotone one and allows us to
detect the presence of synchronization and estimate the boundaries of
the synchronization tongue.

\section{Synchronization of slow oscillations in blood pressure by
respiration from the data of heart rate variability}
\label{Sct:Exp3_Physiology}

In this section we investigate synchronization between the
respiration and rhythmic process of slow regulation of blood
pressure and heart rate from the analysis of univariate data in the
form of the heartbeat time series. \newcorrect{This kind of
synchronization has been experimentally studied in
\cite{Prokhorov:2003_HumanSynchroPRE, Hramov:2006_Prosachivanie,
Janson:2001_PRL, Janson:2002_PRE}.} We studied eight healthy
volunteers. The signal of ECG was recorded with the sampling
frequency 250\,Hz and 16-bit resolution. \correct{ Note, that
according to \cite{Circulation:1996} the sampling frequency 250 Hz
used in our experiments suffices to detect accurately the time
moment of R peak appearance. The experiments were carried out under
paced respiration with the breathing frequency linearly increasing
from 0.05\,Hz to 0.3\,Hz within 30\,min. We specially included the
lower frequencies for paced respiration in order to illustrate the
presence of the most pronounced regime of 1:1 synchronization
between the respiration and slow oscillations in blood pressure.}
The rate of respiration was set by sound pulses. The detailed
description of the experiment is given in
Ref.~\cite{Prokhorov:2003_HumanSynchroPRE}.

Extracting from the ECG signal the sequence of R--R intervals,
i.e., the series of the time intervals between the two successive
R peaks, we obtain the information about the heart rate
variability. The proposed method of detecting synchronization from
uniform data was applied to the sequences of R--R intervals.

\begin{figure}
 \centerline{\scalebox{0.4}{\includegraphics{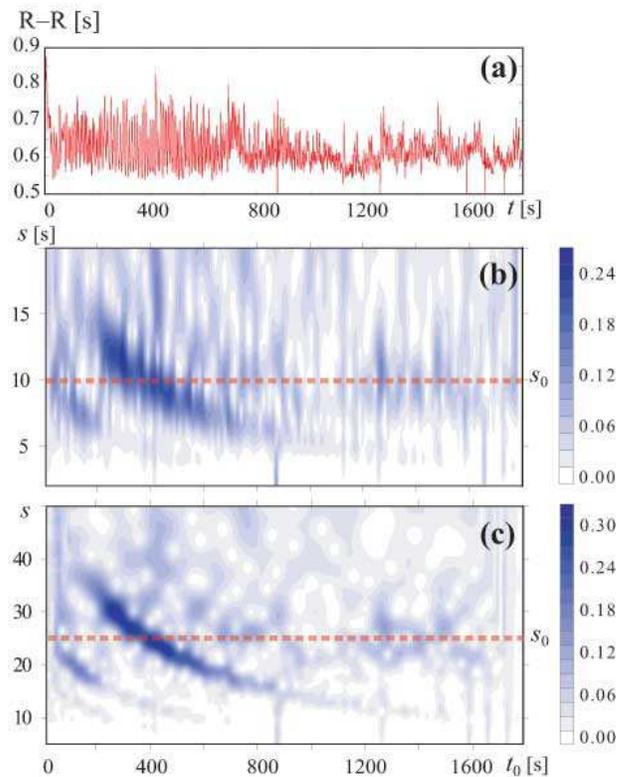}}}
\caption{(Color online) Typical sequence of R--R intervals for the
case of breathing with linearly increasing frequency~(a) and its
wavelet power spectra at $\sigma=2\pi$~(b) and $\sigma=16$~(c). The
dashed lines indicate the time scale $s_0$ corresponding to the
basic frequency $f_0=0.1$\,Hz of slow oscillations in blood
pressure.
 \label{fig8}}
\end{figure}

A typical time series of R--R intervals for breathing at linearly
increasing frequency is shown in Fig.~\ref{fig8}a. Since the
sequence of R--R intervals is not equidistant, we exploit the
technique for applying the continuous wavelet transform to
nonequidistant data. The wavelet spectra $|W(s,t_0)|$ for different
parameters $\sigma$ of the Morlet wavelet are shown in
Figs.~\ref{fig8}b and \ref{fig8}c for the sequence of R--R intervals
presented in Fig.~\ref{fig8}a. For greater $\sigma$ values the
wavelet transform provides higher resolution of frequency
\cite{alkor:2003_WVTBookEng} and better identification of the
dynamics at the time scales corresponding to the basic frequency of
oscillations and the varying respiratory frequency. In the case of
$\sigma=2\pi$ the time scale $s$ of the wavelet transform is very
close to the period $T$ of the Fourier transform and the values of
$s$ are given in seconds in Fig.~\ref{fig8}b. Generally, the time
scale $s$ is related to the frequency $f$ of the Fourier transform
by the following equation:
\begin{equation}
 s=\frac{\sigma+\sqrt{\sigma^2+2}}{4\pi f}.
\label{eq:S_F_SIGMA}
\end{equation}
Because of this, the units on the ordinates are different in
Figs.~\ref{fig8}b and \ref{fig8}c. The wavelet spectra in these
figures demonstrate the high-amplitude component corresponding to
the varying respiratory frequency manifesting itself in the HRV
data. The self-sustained slow oscillations in blood pressure (Mayer
wave) have in humans the basic frequency of about 0.1\,Hz, or
respectively, the basic period close to 10\,s. The power of this
rhythm in the HRV data is less than the power of respiratory
oscillations. As the result, the time scale $s_0$ is weakly
pronounced in the spectra.

\begin{figure}
\centerline{\scalebox{0.4}{\includegraphics{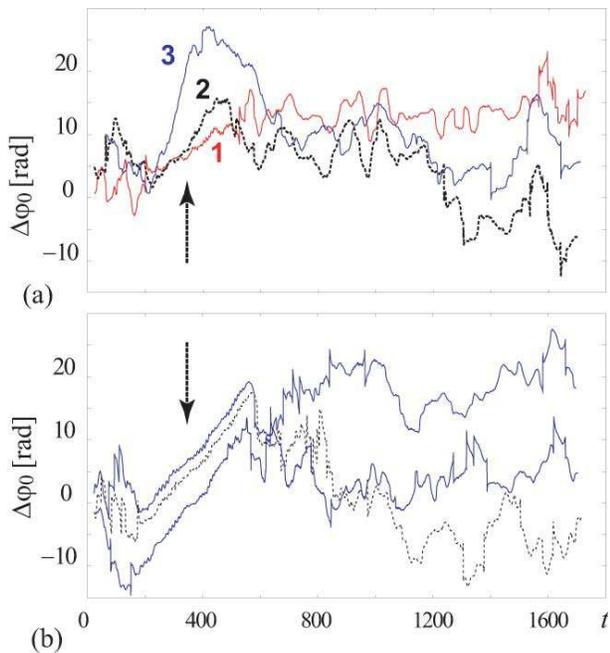}}}
\caption{(Color online) Phase differences $\Delta\varphi_0(t)$
calculated at the time scale $s_0$ corresponding to the basic
frequency $f_0=0.1$\,Hz of the Mayer wave. (a)~Phase differences
computed at different time shifts $\tau$ for R--R intervals of one
of the subjects. The curve numbers correspond to different time
shifts: (1)~$\tau=30$\,s, (2)~$\tau=50$\,s, (3)~$\tau=100$\,s.
(b)~Phase differences computed for R--R intervals of the other three
subjects. \label{fig9}}
\end{figure}

Fig.~\ref{fig9} presents the phase differences $\Delta\varphi_0(t)$
calculated for R--R intervals of four subjects under respiration
with linearly increasing frequency. All the curves in the figure
exhibit the regions with almost linear in the average  variation of
$\Delta\varphi_0(t)$ indicating the presence of synchronous
dynamics. In particular, the region of $1:1$ synchronization is
observed within the interval 200--600\,s when the frequency of
respiration is close to the basic frequency of the Mayer wave. This
region is marked by arrow. \newcorrect{In this region the frequency
of blood pressure slow oscillations is locked by the increasing
frequency of respiration and increases from 0.07 Hz to 0.14 Hz.}
Outside the interval of synchronization, $t<200$\,s and $t>600$\,s,
the phase differences demonstrate fluctuations caused by the high
level of noise and nonstationarity of the experimental data. Some of
these fluctuations take place around an average value as well as in
the case of the driven van der Pol oscillator affected by noise (see
Fig.~\ref{fig3}). \newcorrect{The frequency of blood pressure slow
oscillations demonstrates small fluctuations around the mean value
of about 0.1 Hz outside the interval of synchronization.}

The phase differences in Fig.~\ref{fig9}a are plotted for different
$\tau$. As the time shift $\tau$ increases, so does the range of
$\Delta\varphi_0(t)$ monotone variation in the region of
synchronization. This result agrees well with the results presented
in Sec.~III. \correct{Similar behavior of $\Delta\varphi_0(t)$ is
observed for each of the eight subjects studied. In
Fig.~\ref{fig9}(b) phase differences $\Delta\varphi_0(t)$ computed
for R-R intervals of another three subjects are presented.
\newcorrect{The phase differences demonstrate the wide regions of almost
linear variation for all the subjects. Such behavior of the
considered phase difference cannot be observed in the absence of
synchronization, if only the modulation of blood pressure
oscillations by respiration is present.} These results allow us to
confirm the conclusion that the slow oscillations in blood pressure
can be synchronized by respiration. However, to come to this
conclusion, the proposed method needs only univariate data in
distinction to the methods \cite{Prokhorov:2003_HumanSynchroPRE,
Hramov:2006_Prosachivanie} based on the analysis of bivariate data.
Note, that paper \cite{Prokhorov:2003_HumanSynchroPRE} contains the
more detailed investigation of synchronization between the
respiration and slow oscillations in blood pressure than the present
one.} \newcorrect{Recent reports (see, for examples,
\cite{Rosenblum:1998_Nature, Suder:1998_AJP, Kotani:2000_MIM})
focused on examining the relationship between respiration and heart
rate have shown that there is nonlinear coupling between respiration
and heart rate. In particular, such coupling is well studied for the
respiratory modulation of heart rate \cite{Bishop:1981_AJP,
Kotani:2000_MIM} known as respiratory sinus arrhythmia. The presence
of coupling between the cardiac and respiratory oscillatory
processes has been revealed also using bispectral analysis in
\cite{Jamsek:2003_PRE, Jamsek:2004_PMB} under both spontaneous and
paced respiration. Our results are in agreement with those when
synchronization between the oscillating processes occurs as the
result of their interaction.}

\section{Conclusion}

We have proposed the method for detecting synchronization from
univariate data. The method allows one to detect the presence of
synchronization of the self-sustained oscillator by external force
with varying frequency. To implement the method one needs to
analyze the difference between the oscillator instantaneous phases
calculated at time moments shifted by a certain constant value
with respect to each other. The instantaneous phases are defined
at the oscillator basic frequency using continuous wavelet
transform with the Morlet wavelet as the mother wavelet function.
The necessary condition for the method application is the
variation of the frequency of the driving signal. The method
efficiency is illustrated using both numerical and experimental
univariate data under sufficiently high levels of noise and
inaccuracy of the basic time scale definition.

We applied the proposed method to studying synchronization between
the respiration and slow oscillations in blood pressure from
univariate data in the form of R--R intervals. The presence of
synchronization between these rhythmic processes is demonstrated
within the wide time interval. The knowledge about synchronization
between the rhythms of the cardiovascular system under paced
respiration is useful for the diagnostics of its state
\cite{Ancona:2005_Physiological Measurement}. \correct{The method
allows one to detect the presence of synchronization from the
analysis of the data of Holter monitor widely used in cardiology.}

The proposed method can be used for the analysis of synchronization
even in the case when the law of the driving frequency variation is
unknown. If the frequency of the external driving varies in the wide
range, the analysis of the oscillator response to the unknown driving
force allows one to make a conclusion about the presence or absence of
synchronization in the system under investigation.

\section*{Acknowledgments}
\label{Sct: Acknowledgments}

We thank Dr. Svetlana Eremina for English language support. This
work is supported by the Russian Foundation for Basic Research,
Grants 05--02--16273, 07--02--00044, 07--02--00747 and
07--02--00589, and the President Program for support of the leading
scientific schools in the Russian Federation, Grant
No.~SCH-4167.2006.2. A.E.H. acknowledges support from CRDF, Grant
No.~Y2--P--06--06. A.E.H. and A.A.K. thank the ``Dynasty''
Foundation for the financial support.


\begin{thebibliography}{32}
\expandafter\ifx\csname
natexlab\endcsname\relax\def\natexlab#1{#1}\fi
\expandafter\ifx\csname bibnamefont\endcsname\relax
  \def\bibnamefont#1{#1}\fi
\expandafter\ifx\csname bibfnamefont\endcsname\relax
  \def\bibfnamefont#1{#1}\fi
\expandafter\ifx\csname citenamefont\endcsname\relax
  \def\citenamefont#1{#1}\fi
\expandafter\ifx\csname url\endcsname\relax
  \def\url#1{\texttt{#1}}\fi
\expandafter\ifx\csname urlprefix\endcsname\relax\def\urlprefix{URL
}\fi \providecommand{\bibinfo}[2]{#2}
\providecommand{\eprint}[2][]{\url{#2}}

\bibitem[{\citenamefont{{Blekhman I.I.}}(1971)}]{Blekhman:1971_SynhroBookEngl}
\bibinfo{author}{\bibnamefont{{Blekhman I.I.}}},
  \emph{\bibinfo{title}{Synchronization of dynamical systems}}
  (\bibinfo{publisher}{Moscow, Nauka}, \bibinfo{year}{1971}).

\bibitem[{\citenamefont{{Blekhman I.I.}}(1988)}]{Blekhman:1988_SynchroBook}
\bibinfo{author}{\bibnamefont{{Blekhman I.I.}}},
  \emph{\bibinfo{title}{Synchronization in Science and Technology}}
  (\bibinfo{publisher}{American Society of Mechanical Engineers},
  \bibinfo{year}{1988}).

\bibitem[{\citenamefont{{Pikovsky~A., Rosenblum~M.,
  Kurths~J.}}(2001)}]{Pikovsky:2002_SynhroBook}
\bibinfo{author}{\bibnamefont{{Pikovsky~A., Rosenblum~M., Kurths~J.}}},
  \emph{\bibinfo{title}{Synhronization: a universal concept in nonlinear
  sciences}} (\bibinfo{publisher}{Cambridge University Press, Cambridge},
  \bibinfo{year}{2001}).

\bibitem[{\citenamefont{{Boccaletti S., Kurths J., Osipov G., Valladares D.L.,
  Zhou C.}}(2002)}]{Boccaletti:2002_ChaosSynchro}
\bibinfo{author}{\bibnamefont{{Boccaletti S., Kurths J., Osipov G., Valladares
  D.L., Zhou C.}}}, \bibinfo{journal}{Physics Reports}
  \textbf{\bibinfo{volume}{366}}, \bibinfo{pages}{1} (\bibinfo{year}{2002}).

\bibitem[{\citenamefont{{Rosenblum M., Pikovsky A., Kurths J., Schafer C., Tass
  P.}}(2001)}]{Rosenblum:2001_HandbookBiologicalPhysics}
\bibinfo{author}{\bibnamefont{{Rosenblum M., Pikovsky A., Kurths J., Schafer
  C., Tass P.}}}, in \emph{\bibinfo{booktitle}{Handbook of Biological Physics}}
  (\bibinfo{publisher}{Elsiver Science}, \bibinfo{year}{2001}), pp.
  \bibinfo{pages}{279--321}.

\bibitem[{\citenamefont{{Meinecke F.C., Ziehe A., Kurths J., M\"uller
  K.-R.}}(2005)}]{Meinecke:2005_Prosachivanie}
\bibinfo{author}{\bibnamefont{{Meinecke F.C., Ziehe A., Kurths J., M\"uller
  K.-R.}}}, \bibinfo{journal}{Phys. Rev. Lett.} \textbf{\bibinfo{volume}{94}},
  \bibinfo{pages}{084102} (\bibinfo{year}{2005}).

\bibitem[{\citenamefont{{Hramov A.E., Koronovskii
  A.A.}}(2004)}]{Hramov:2004_Chaos}
\bibinfo{author}{\bibnamefont{{Hramov A.E., Koronovskii A.A.}}},
  \bibinfo{journal}{Chaos} \textbf{\bibinfo{volume}{14}}, \bibinfo{pages}{603}
  (\bibinfo{year}{2004}).

\bibitem[{\citenamefont{{Hramov A.E., Koronovskii A.A., Kurovskaya M.K.,
  Moskalenko O.I.}}(2005)}]{Aeh:2005_SpectralComponents}
\bibinfo{author}{\bibnamefont{{Hramov A.E., Koronovskii A.A., Kurovskaya M.K.,
  Moskalenko O.I.}}}, \bibinfo{journal}{Phys. Rev. E}
  \textbf{\bibinfo{volume}{71}}, \bibinfo{pages}{056204}
  (\bibinfo{year}{2005}).

\bibitem[{\citenamefont{{Pecora L.M.,
  Carroll T.L.}}(1990)}]{Pecora:1990_ChaosSynchro}
\bibinfo{author}{\bibnamefont{{Pecora~L.M., Carroll~T.L.}}},
  \bibinfo{journal}{Phys. Rev. Lett.} \textbf{\bibinfo{volume}{64}},
  \bibinfo{pages}{821} (\bibinfo{year}{1990}).

\bibitem[{\citenamefont{{Pecora L.M., Carroll T.L., Jonson G.A., Mar
  D.J.}}(1997)}]{Pecora:1997_SynchroChaos}
\bibinfo{author}{\bibnamefont{{Pecora L.M., Carroll T.L., Jonson G.A., Mar
  D.J.}}}, \bibinfo{journal}{Chaos} \textbf{\bibinfo{volume}{7}},
  \bibinfo{pages}{520} (\bibinfo{year}{1997}).

\bibitem[{\citenamefont{{Pikovsky A., Rosenblum M.,
  Kurths J.}}(2000)}]{Pikovsky:2000_SynchroReview}
\bibinfo{author}{\bibnamefont{{Pikovsky A., Rosenblum M., Kurths J.}}},
  \bibinfo{journal}{Int. J. Bifurcation and Chaos}
  \textbf{\bibinfo{volume}{10}}, \bibinfo{pages}{2291} (\bibinfo{year}{2000}).

\bibitem[{\citenamefont{{Boccaletti S., Pecora L.M.,
  Pelaez A.}}(2001)}]{Boccaletti:2001_UnifingSynchro}
\bibinfo{author}{\bibnamefont{{Boccaletti S., Pecora L.M., Pelaez A.}}},
  \bibinfo{journal}{Phys. Rev. E} \textbf{\bibinfo{volume}{63}},
  \bibinfo{pages}{066219} (\bibinfo{year}{2001}).

\bibitem[{\citenamefont{{Rulkov N.F., Sushchik M.M., Tsimring L.S.,
  Abarbanel H.D.I.}}(1995)}]{Rulkov:1995_GeneralSynchro}
\bibinfo{author}{\bibnamefont{{Rulkov N.F., Sushchik M.M., Tsimring L.S.,
  Abarbanel H.D.I.}}}, \bibinfo{journal}{Phys. Rev. E}
  \textbf{\bibinfo{volume}{51}}, \bibinfo{pages}{980} (\bibinfo{year}{1995}).

\bibitem[{\citenamefont{{Pyragas K.}}(1996)}]{Pyragas:1996_WeakAndStrongSynchr%
o}
\bibinfo{author}{\bibnamefont{{Pyragas K.}}}, \bibinfo{journal}{Phys. Rev. E}
  \textbf{\bibinfo{volume}{54}}, \bibinfo{pages}{R4508} (\bibinfo{year}{1996}).

\bibitem[{\citenamefont{Tass et~al.}(1998)\citenamefont{Tass, Rosenblum, Weule,
  Kurths, Pikovsky, Volkmann, Schnitzler, and Freund}}]{Tass:1998_NeuroSynchro}
\bibinfo{author}{\bibfnamefont{P.}~\bibnamefont{Tass}},
  \bibinfo{author}{\bibfnamefont{M.~G.} \bibnamefont{Rosenblum}},
  \bibinfo{author}{\bibfnamefont{J.}~\bibnamefont{Weule}},
  \bibinfo{author}{\bibfnamefont{J.}~\bibnamefont{Kurths}},
  \bibinfo{author}{\bibfnamefont{A.}~\bibnamefont{Pikovsky}},
  \bibinfo{author}{\bibfnamefont{J.}~\bibnamefont{Volkmann}},
  \bibinfo{author}{\bibfnamefont{A.}~\bibnamefont{Schnitzler}},
  \bibnamefont{and} \bibinfo{author}{\bibfnamefont{H.-J.}
  \bibnamefont{Freund}}, \bibinfo{journal}{Phys. Rev. Lett.}
  \textbf{\bibinfo{volume}{81}}, \bibinfo{pages}{3291} (\bibinfo{year}{1998}).

\bibitem[{\citenamefont{Tass et~al.}(2003)\citenamefont{Tass, Fieseler,
  Dammers, Dolan, Morosan, Majtanik, Boers, Muren, Zilles, and
  Fink}}]{Tass:2003_NeuroSynchro}
\bibinfo{author}{\bibfnamefont{P.~A.} \bibnamefont{Tass}},
  \bibinfo{author}{\bibfnamefont{T.}~\bibnamefont{Fieseler}},
  \bibinfo{author}{\bibfnamefont{J.}~\bibnamefont{Dammers}},
  \bibinfo{author}{\bibfnamefont{K.}~\bibnamefont{Dolan}},
  \bibinfo{author}{\bibfnamefont{P.}~\bibnamefont{Morosan}},
  \bibinfo{author}{\bibfnamefont{M.}~\bibnamefont{Majtanik}},
  \bibinfo{author}{\bibfnamefont{F.}~\bibnamefont{Boers}},
  \bibinfo{author}{\bibfnamefont{A.}~\bibnamefont{Muren}},
  \bibinfo{author}{\bibfnamefont{K.}~\bibnamefont{Zilles}}, \bibnamefont{and}
  \bibinfo{author}{\bibfnamefont{G.~R.} \bibnamefont{Fink}},
  \bibinfo{journal}{Phys. Rev. Lett.} \textbf{\bibinfo{volume}{90}},
  \bibinfo{pages}{088101} (\bibinfo{year}{2003}).

\bibitem[{\citenamefont{{Chavez M., Adam C., Navarro, Boccaletti S., Martinerie
  J.}}(2005)}]{Boccaletti:2005_Chaos_TSS}
\bibinfo{author}{\bibnamefont{{Chavez M., Adam C., Navarro, Boccaletti S.,
  Martinerie J.}}}, \bibinfo{journal}{Chaos} \textbf{\bibinfo{volume}{15}}
  (\bibinfo{year}{2005}).



\bibitem[{\citenamefont{{Sch\"afer C., Rosenblum M.G., Abel H.-H., Kurths
  J.}}(1999)}]{Schafer:1999_cardio}
\bibinfo{author}{\bibnamefont{{Sch\"afer C., Rosenblum M.G., Abel H.-H., Kurths
  J.}}}, \bibinfo{journal}{Phys. Rev. E} \textbf{\bibinfo{volume}{60}},
  \bibinfo{pages}{857} (\bibinfo{year}{1999}).

\bibitem[{\citenamefont{{Bra\v{c}i\v{c}-Lotri\v{c} M., Stefanovska
  A.}}(2000)}]{Stefanovska:2000_cardio_Physica_A}
\bibinfo{author}{\bibnamefont{{Bra\v{c}i\v{c}-Lotri\v{c} M., Stefanovska A.}}},
  \bibinfo{journal}{Physica A} \textbf{\bibinfo{volume}{283}},
  \bibinfo{pages}{451} (\bibinfo{year}{2000}).

\bibitem[{\citenamefont{{Rzeczinski S., Janson N.B., Balanov A.G., McClintock
  P.V.E.}}(2002)}]{Rzeczinski:2002_cardio}
\bibinfo{author}{\bibnamefont{{Rzeczinski S., Janson N.B., Balanov A.G.,
  McClintock P.V.E.}}}, \bibinfo{journal}{Phys. Rev. E}
  \textbf{\bibinfo{volume}{66}}, \bibinfo{pages}{051909}
  (\bibinfo{year}{2002}).

\bibitem[{\citenamefont{{Prokhorov M.D., Ponomarenko V.I., Gridnev V.I., Bodrov
  M.B., Bespyatov A.B.}}(2003)}]{Prokhorov:2003_HumanSynchroPRE}
\bibinfo{author}{\bibnamefont{{Prokhorov M.D., Ponomarenko V.I., Gridnev V.I.,
  Bodrov M.B., Bespyatov A.B.}}}, \bibinfo{journal}{Phys. Rev. E}
  \textbf{\bibinfo{volume}{68}}, \bibinfo{pages}{041913}
  (\bibinfo{year}{2003}).

\bibitem[{\citenamefont{{Hramov A.E., Koronovskii A.A., Ponomarenko V.I.,
  Prokhorov M.D.}}(2006)}]{Hramov:2006_Prosachivanie}
\bibinfo{author}{\bibnamefont{{Hramov A.E., Koronovskii A.A., Ponomarenko V.I.,
  Prokhorov M.D.}}}, \bibinfo{journal}{Phys. Rev. E}
  \textbf{\bibinfo{volume}{73}}, \bibinfo{pages}{026208}
  (\bibinfo{year}{2006}).

\bibitem[{\citenamefont{{Hramov A.E., Koronovskii A.A., Levin
  Yu.I}}(2005)}]{Hramov:2005_JETP}
\bibinfo{author}{\bibnamefont{{Hramov A.E., Koronovskii A.A., Levin Yu.I}}},
  \bibinfo{journal}{JETP} \textbf{\bibinfo{volume}{127}}, \bibinfo{pages}{886}
  (\bibinfo{year}{2005}).

\bibitem[{\citenamefont{{Hramov A.E., Koronovskii
  A.A.}}(2005)}]{Aeh:2005_TSS:PhysicaD}
\bibinfo{author}{\bibnamefont{{Hramov A.E., Koronovskii A.A.}}},
  \bibinfo{journal}{Physica D} \textbf{\bibinfo{volume}{206}},
  \bibinfo{pages}{252} (\bibinfo{year}{2005}).

\bibitem[{\citenamefont{{Hramov A.E., Koronovskii A.A., Popov P.V., Rempen I.S.}}(2005)}]{Hramov:2005_Chaos_BWO}
\bibinfo{author}{\bibnamefont{{Hramov A.E., Koronovskii A.A., Popov P.V., Rempen I.S.}}}, \bibinfo{journal}{Chaos}
\textbf{\bibinfo{volume}{15}}, \bibinfo{pages}{013705}
(\bibinfo{year}{2005}).



\bibitem[{\citenamefont{{Malpas S.}}(2002)}]{Malpas:2002_cardio}
\bibinfo{author}{\bibnamefont{{Malpas S.}}}, \bibinfo{journal}{Am. J. Physiol.
  Heart Circ. Physiol.} \textbf{\bibinfo{volume}{282}}, \bibinfo{pages}{H6}
  (\bibinfo{year}{2002}).

\bibitem[{\citenamefont{{Stefanovska A., Ho\v{z}i\v{c}
  M.}}(2000)}]{Stefanovska:2000_cardio}
\bibinfo{author}{\bibnamefont{{Stefanovska A., Ho\v{z}i\v{c} M.}}},
  \bibinfo{journal}{Prog. Theor. Phys. Suppl.} 139,
  270 (\bibinfo{year}{2000}).

\bibitem[{\citenamefont{{Adler R.}}(1947)}]{Adler:1949}
\bibinfo{author}{\bibnamefont{{Adler R.}}}, \bibinfo{journal}{Proc. IRE}
  \textbf{\bibinfo{volume}{35}}, \bibinfo{pages}{1415--1423} (\bibinfo{year}{1947}).

\bibitem[{\citenamefont{{Koronovskii A.A., Hramov
  A.E.}}(2004)}]{Koronovskii:2004_JETPLettersEngl}
\bibinfo{author}{\bibnamefont{{Koronovskii A.A., Hramov A.E.}}},
  \bibinfo{journal}{JETP Lett.} \textbf{\bibinfo{volume}{79}},
  \bibinfo{pages}{316} (\bibinfo{year}{2004}).

\bibitem[{Wav(2004)}]{WaveletsInPhysics:2004}
\emph{\bibinfo{title}{Wavelets in Physics}}
(\bibinfo{publisher}{Cambridge
  University Press, Cambridge}, \bibinfo{year}{2004}), \bibinfo{edition}{{J.C.
  Van den Berg}} ed.

\bibitem[{\citenamefont{{Koronovskii A.A.,
  Hramov A.E.}}(2003)}]{alkor:2003_WVTBookEng}
\bibinfo{author}{\bibnamefont{{Koronovskii A.A., Hramov A.E.}}},
  \emph{\bibinfo{title}{Continuous wavelet analysis and its applications}}
  (\bibinfo{publisher}{Moscow, Fizmatlit}, \bibinfo{year}{2003}).



{

\bibitem{Lachaux:1999}
Lachaux J.P., Rodriguez E., Martinerie J., Varela F.J., Human Brain
Mapping \textbf{8}, 194, (1999).

\bibitem{Lachaux:2000}
Lachaux J.P., Rodriguez E., Van Quyen M.L., Lutz A., Martinerie J.,
Varela F.J., Int. J. Bifurcation and Chaos. \textbf{10}, 2429
(2000).

\bibitem{Lachaux:2001}
Le Van Quyen M., Foucher J., Lachaux J.P., Rodriguez E., Lutz A.,
Martinerie J., Varela F.J. J. Neuroscience Methods \textbf{111}, 83
(2001).

\bibitem{Lachaux:2002_BrainCoherence}
Lachaux J.P., Lutz A., Rudrauf D., Cosmelli D., Le Van Quyen M.,
Martinerie J., Varela F. {Neurophysiol. Clin.}, \textbf{32}, 157
(2002).

\bibitem{Quyen:2001_HTvsWVT}
 Quyen M.~L., Foucher J.,  Lachaux J.-P., Rodriguez E., Lutz A.,
Martinerie J.,  Varela F.~J. J. Neuroscience Methods \textbf{111},
83 (2001).

\bibitem{Quiroga:2002}
Quiroga R.Q., Kraskov A., Kreuz T., Grassberger P. Phys. Rev. E
\textbf{65}, 041903 (2002).



\bibitem{Turalska:2005}
Turalska M., Kolodziej W., Latka M., Latka D., West B.J.  Shock
\textbf{23}, Suppl. 3, 90 (2005).

\bibitem{DeShazer:2001_WVT_LaserArray}
DeShazer D.~J., Breban R., Ott E., and Roy R. Phys. Rev. Lett.
\textbf{87}, 044101 (2001).

\bibitem{Sosnovtseva:2002_Wvt}
 Sosnovtseva O.~V.,  Pavlov A.~N., Mosekilde E., and
Holstein-Rathlou N.-H. {Phys. Rev. E} \textbf{66}, 061909 (2002).


\bibitem{BANDRIVSKYY:2004}
Bandrivskyy A., Bernjak A.,  McClintock P. V. E., and Stefanovska A.
Cardiovasc. Eng. Int. J. \textbf{4}, 89 (2004).}








\bibitem[{\citenamefont{{Grossman A. and
  Morlet J.}}(1984)}]{Grossman:1984_Morlet}
\bibinfo{author}{\bibnamefont{{Grossman A. and Morlet J.}}},
  \bibinfo{journal}{SIAM J. Math. Anal.} \textbf{\bibinfo{volume}{15}},
  \bibinfo{pages}{273} (\bibinfo{year}{1984}).

\bibitem[{\citenamefont{{Press W.H., Teukolsky S.A., Vetterling W.T., Flannery
  B.T.}}(1997)}]{NumericalRecipes:1997}
\bibinfo{author}{\bibnamefont{{Press W.H., Teukolsky S.A., Vetterling W.T.,
  Flannery B.T.}}}, \emph{\bibinfo{title}{Numerical Recipes}}
  (\bibinfo{publisher}{Cambridge University Press, Cambridge},
  \bibinfo{year}{1997}).



\bibitem{Janson:2001_PRL} Janson N. B., Balanov A. G., Anishchenko V. S., McClintock P. V.
E. Phys. Rev. E \textbf{86}, 1749 (2001).


\bibitem{Janson:2002_PRE} Janson N. B., Balanov A. G., Anishchenko V. S., McClintock P. V.
E. Phys. Rev. E \textbf{65}, 036212 (2002).


\bibitem{Circulation:1996}
Task Force of the ESC and NASPE, Circulation \textbf{93}, 1043,
(1996).

\bibitem{Rosenblum:1998_Nature} Sh\"afer C., Rosenblum M.G., Abel H.H. Nature \textbf{392}, 239 (1998).

\bibitem{Suder:1998_AJP} Suder K., Drepper F.R., Schiek M., Abel H.H. Am. J. Physiol. \textbf{275}, H33 (1998).


\bibitem{Bishop:1981_AJP} Hirsch J.A., Bishop B. Am. J. Physiol. \textbf{241}, H620 (1981).

\bibitem{Kotani:2000_MIM} Kotani K., Hidaka I., Yamamoto Y., Ozono S. Method Inform Med. \textbf{39}, 153 (2000).

\bibitem{Jamsek:2003_PRE} Jam\v{s}ek J., Stefanovska A., McClintock P.V.E.,
Khovanov I.A. Phys. Rev. E \textbf{68}, 016201 (2003).

\bibitem{Jamsek:2004_PMB} Jam\v{s}ek J., Stefanovska A., McClintock P.V.E. Phys. Med. Biol. \textbf{49}, 4407
(2004).


\bibitem[{\citenamefont{{N. Ancona, R. Maestri, D.
Marinazzo, L. Nitti, M. Pellicoro, G.D. Pinna, S.
Stramaglia}}(2005)}]{Ancona:2005_Physiological Measurement}
\bibinfo{author}{\bibnamefont{{N. Ancona, R. Maestri, D.
Marinazzo, L. Nitti, M. Pellicoro, G.D. Pinna, S. Stramaglia}}},
\bibinfo{journal}{Physiological Measurement} \textbf{\bibinfo{volume}{26}},
  \bibinfo{pages}{363--372} (\bibinfo{year}{2005}).


\end{thebibliography}

\end{document}